# A New Paradigm to Half-metallicity in Graphene Nanoribbons

Jin Yu and Wanlin Guo[*]

*Key Laboratory for Intelligent Nano Materials and Devices of MOE and State Key Laboratory of Mechanics and Control of Mechanical Structures, Institute of Nano Science, Nanjing University of Aeronautics and Astronautics, Nanjing, 210016, China*

ABSTRACT: In contrast to the well recognized transverse-electric-field-induced half-metallicity in zigzag graphene nanoribbons, here we demonstrate by first-principles calculations that zigzag graphene nanoribbons sandwiched between hexagonal boron nitride nanoribbons or sheets can be tuned into half-metal simply by a bias voltage or a moderate compressive strain. The half-metallicity is attributed to an enhanced coupling effect of spontaneous polarization and asymmetrical exchange correlation along the ribbon width. The findings should open a viable route for efficient spin-resolved band engineering in graphene based devices that are compatible with the current technology of semiconductor industry.

TOC graphic

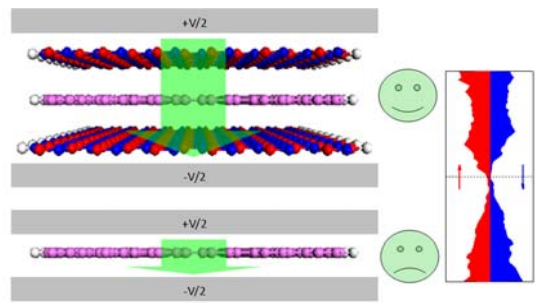

Graphene nanoribbons (GNRs) are attractive for the next-generation electronic devices due to their energy gap induced by quantum confinement effect,[1,2] high electron mobility[3,4] and excellent thermal conductance.[5-7] Particularly, it's predicted that the zigzag-edged GNRs (ZGNRs) exhibit amazing edge ferromagnetism[1,8-10] and can be further developed into a half-metal upon a sufficient transversely applied electric potential difference across the ribbons.[11-13] However, in contrast to the widely applied gate voltage or bias electric fields in the current technology, the required transverse potential difference for half-metal transition increases with increasing ribbon width and its realization is challenging. To circumvent this difficulty, many alternative routes have been proposed theoretically to provide an equivalent potential difference across the ribbon width, such as being chemically functionalized or embedded in boron nitride (BN) sheet to form hybrid systems.[14-16] It is especially recognized that hexagonal boron nitride sheet can serve as a best matching substrate for graphene to realize its high carrier mobility.[17,18] First-principles calculations also showed that a band gap can be opened in graphene sandwiched between BN sheets.[19-22] However, none of these alternative routes can be effective enough to induce half-metallic transition. In this letter, we reveal by extensive first-principles calculations that a ZGNR sandwiched between two BN nanoribbons (BNNRs) or sheets can be easily turned into semi-metallic state by bias voltage or a moderate perpendicular pressure, as the inherent transverse polarization of the sandwiched structure can be significantly enhanced by bias voltage or reduced interlayer spacing. The results show a new and efficient route compatible with the current technology for realizing half-metallicity in graphene based systems.

A schematic illustration of the model geometry used in the calculations is presented in Figure 1. The ZGNR is sandwiched between two BNNRs (Z-BN-G-BNNRs) with all the atoms being in AA stacked. Both the top and bottom BNNRs are patterned in the same orientation with the boron-ended edge at left and nitride-ended edge at right. Consequently, the carbon atom at the left edge of the sandwich is denoted as $C_B$, and that at the right edge as $C_N$. It is obvious that the $C_B$ and $C_N$ belong to different sublattice in the ZGNR. Here, we assume the $C_B$ is at the A-sublattice, and the $C_N$ is at the B-sublattice. Following previous conventions, the Z-BN-G-BNNRs are classified by the number of zigzag chains $N_z$ and denoted as $N_z$-Z-BN-G-BNNRs. A one-dimensional periodic boundary condition is applied along the ribbon edge. Both of the vacuum distances between ribbon edges and between sandwiches of two adjacent images are set to be at least 1.2 nm. All the atoms at the ribbon edges are passivated with hydrogen atoms.

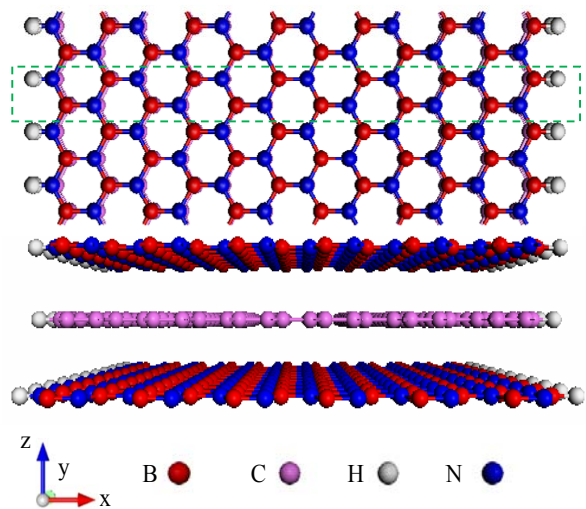

**Figure 1.** Top panel: Top view of atom structures of the sandwiched $N_z$-Z-BN-G-BNNR. The green dashed frame shows the unit cells taken into account in our computations. Bottom panel: Front view of the $N_z$-BN-G-BNNR. A width of $N_z$=12 is used for illustration here.

To confirm the relative energy stability of the heterostructures, we first checked the total free energy of the usual AA, ABC and ABA stacked sandwiches. It is found that the AA stacked 16-Z-BN-G-BNNRs is more favorable in energy, the total free energy of which is 13 meV and 265 meV lower than that of the corresponding ABC and ABA, respectively. Further analyses shown that the electronic and magnetic properties of the sandwiches are only slightly affected by the stacking configurations (see Supporting Information (SI) and Figure S1). So we will mainly discuss the properties of the sandwiches with AA stacked manner. Spin-unpolarized as well as spin-polarized calculations on the electronic properties of the Z-BN-G-BNNRs were performed. We found that the ground state of the sandwiched structure remains to be magnetic and the polarized spin densities are localized at the edges of the ZGNRs as shown in Figure 2a. The spin ordering in the 16-Z-BN-G-BNNR is that the polarized spins are parallel aligned along the ribbon edge and antiferromagnetically coupled between two opposite edges of the constituent ZGNR, akin to the distribution of spin pattern in a single-layer ZGNR. However, by scrutinizing the band structure shown in Figure 2b, remarkable difference can be found in comparison with that of the corresponding single-layer 16-ZGNR. In particular, we find that there is a significant spin splitting near $2\pi/3$ of the Brillouin zone. The band gap of up-spin channel is reduced to 0.10 eV while that of down-spin channel is increased to 0.30 eV from the gap 0.20 eV of pristine 16-ZGNR, rendering the 16-Z-BN-G-BNNR a natural semiconductor without applied electric field or edge chemical modification via sophisticated process. Through extensive wave function analyses, we have identified that the bands around the Fermi level within the energy window from -1 to 1 eV are all from the GNR, and the electronic states from the BNNRs are located either 3 eV higher or 2 eV lower than the Fermi level. It thus can be envisioned that the transport behaviors of such sandwiched nanoribbons will be dominated by the GNR, while the BNNRs serve as shielding layers.

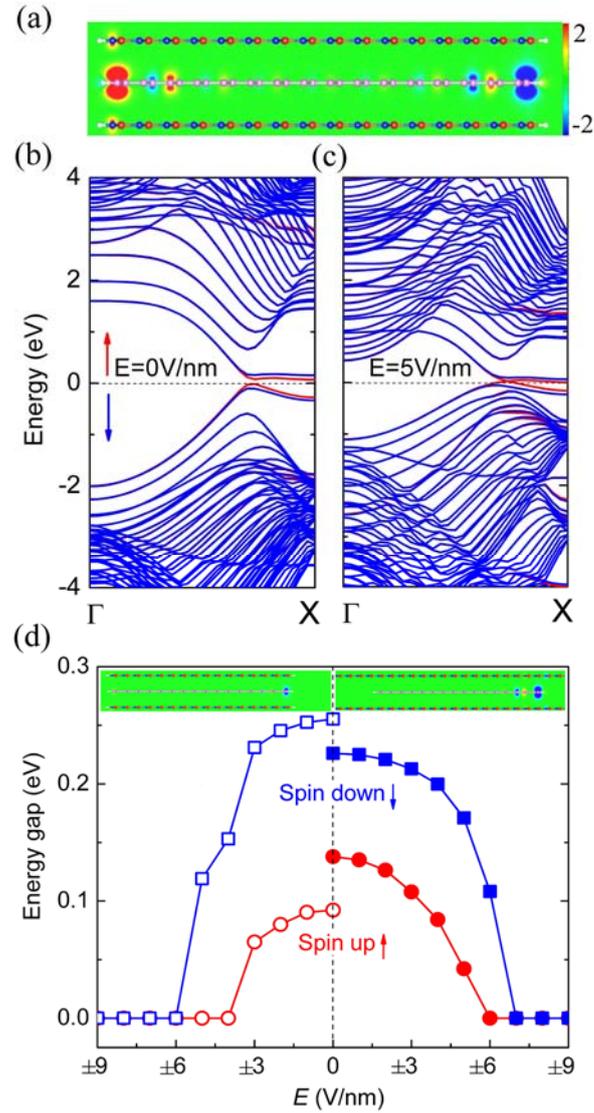

**Figure 2.** (a) Vertical slices of the spin density distribution in a 16-Z-BN-G-BNNR. The isosurface level is set to 1.2 e · nm$^{-3}$. (b) Band structures of the 16-Z-BN-G-BNNR without any fields and (c) under a bias field of $E$ = 5 V/nm, with the Fermi level being set to 0. (d) Bias field induced shift in Energy gap in the 16-ZGNR sandwiched between two BN nanoribbons (left) and BN sheets (right). The left and right insets show the spin density distribution of the ZGNR sandwiched between BN nanoribbons at $E$ = 5 V/nm and BN sheets at $E$ = 6 V/nm, respectively. The red and blue lines in (b,c,d) are for up-spin and down-spin curves, respectively.

It is of practical importance that the energy gap of the semiconducting Z-BN-G-BNNRs can be efficiently modulated by applying a bias voltage. The bias field is simulated by a periodic sawtooth-type potential normal to the ribbon plane and the positive bias is defined pointing from the bottom BNNR to the top BNNR. Figure 2c presents the band structure of the 16-Z-BN-G-BNNR in $E$ = 5 V/nm, at which the system has already turned out to be a half-metal. Further wave function analyses show that the bias voltage has significant effect on the BNNRs as well. Though bands attributed by the BNNRs shift toward the Fermi level more notably than that of the ZGNRs, the spin character and transport properties of the system are dominated by the ZGNRs. This bias voltage induced half-metallicity has distinguished advantage over that induced by transverse fields as proposed in the pioneering work.[11] The spin-dependent energy gaps of the 16-Z-BN-G-BNNR as a function of the bias $E$ from 0 V/nm to ±9 V/nm are shown in the left half of Figure 2d. As the top and bottom layers of BNNR are identical, the negative and positive biases have exactly the same effect. Without bias voltage, the energy gap for down-spin is much larger than that of un-spin mainly due to the intrinsic polarization of the BNNRs. With increasing bias voltage, the gaps for both spins decrease, but the gap for up-spin drops to zero before that for down-spin, leading to a half-metallic state. It is interesting that very similar results can be obtained for a ZGNR sandwiched between two infinite large BN sheets as shown in the right half of the Figure 2d, where AA stacking is assumed between the ZGNR and the BN sheets. For both kinds of sandwiches, the ZGNR becomes a half-metal when the bias voltage is increased to about ±4~ ±6V/nm.

The insets of Figure 2d show the spin density distribution of both sandwiches in the half-metallic state.

To understand the mechanism of the bias voltage induced energy band shift, we then examine the variations of the differential electrostatic potential ($\Delta_P = P_{BN-G-BN} - P_G - P_{2BN}$) of the 16-Z-BN-G-BNNR with increasing bias voltage as shown in the left column of Figure 3. Where, $P_{BN-G-BN}$, $P_G$ and $P_{2BN}$ present the electrostatic potential of a 16-Z-BN-G-BNNR, a freestanding 16-ZGNR and a bilayer suspending 16-ZBNNRs within the same supercell, respectively. Without bias field, the plane-averaged electrostatic potential is significantly reduced near the $C_N$-edge (left) of the sandwich but is enhanced relatively around the $C_B$-edge (right), with slight change in the middle part of the ribbon (Figure 3a). This redistribution in the differential potential is mainly attributed to the different local potentials at B and N atoms and the spontaneous electric polarization in the ZBNNRs. As a result, the corresponding potential shifts of the up-spin and down-spin in the ZGNR are different, making the system a semiconductor with different energy gaps for the up- and down-spin channels. At a bias voltage of $E = 1$ V/nm, the differential potentials near the $C_N$-edge and the $C_B$-edge are far from symmetric as shown in Figure 3b, but become nearly symmetric when $E$ reaches 5 V/nm, Figure 3c. Therefore, a bias voltage can significantly redistribute the differential potentials, leading to shift in its electronic properties.

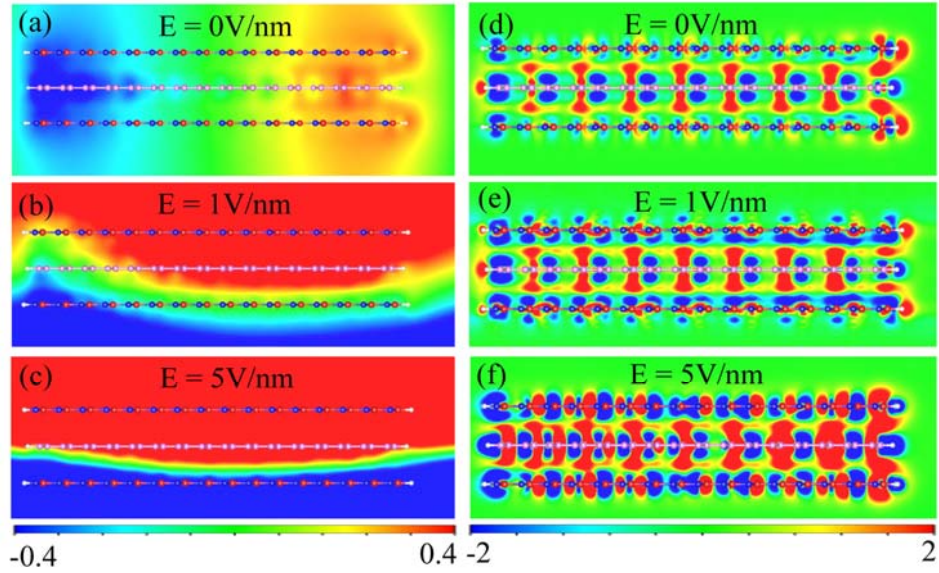

**Figure 3.** Distribution of the differential electrostatic potential of the 16-Z-BN-G-BNNR along the width direction *x* with (a) $E$ = 0 V/nm, (b) 1 V/nm and (c) 5 V/nm. (d,e,f) The corresponding differential charge of the sandwich. The isosurface level is set to 1.5 e·nm$^{-3}$.

To further explore the interaction of the electrons in different ribbon layers, we examine the charge redistribution among the ZGNR and ZBNNRs in the sandwich. This is performed by calculating the charge density difference $\Delta\rho=\rho_{BN-G-BN}-\rho_G-\rho_{2BN}$, where $\rho_{BN-G-BN}$, $\rho_G$ and $\rho_{2BN}$ are the total charge densities of the sandwich, the individual ZGNR and the bilayer ZBNNRs, respectively. Due to the spontaneous electric polarization in the ZBNNRs, remarkable charge transfer occurs from the B-sublattice to the A-sbulattice along the orientation from $C_N$ to $C_B$ in the ZGNR, while very slight charge transfer occurs in the ZBNNR plane (Figure 3d). Between the ZBNNR and ZGNR, there is a sparse charge accumulation, forming a nearly free electron state.[23] With increasing bias voltage, stronger charge redistributions occur in each layer and between the layers as shown by Figure 3e and 3f. Similar mechanism can be found for a ZGNR sandwiched between two BN sheets. These results strongly suggest that the coupling effect between the spontaneous polarization of the sandwiched structure and the applied bias voltage opens an efficient path for spin-resolved band engineering of the ZGNRs.

On the other hand, the electronic structure of few-layer graphene has been shown to be very sensitive to the interlayer spacing,[15] which can be easily reduced by nanoindentation.[24] We expect similar behavior in our sandwiched Z-BN-G-BNNRs. To show the effect of nanoindentation, the interlayer distance between the ZGNR and ZBNNR planes surface is adjusted by fixing the $z$ coordinates of the ZBNNR and ZGNR atoms at given values, while allowing full relaxation of the atoms within their planes. It is shown that the sandwiched nanoribbons can be tuned into half-metal via reducing the interlayer spacing, see Figure 4a. When the interlayer distance $d$ is fixed to its equilibrium value of 0.334 nm, the up- and down-spin gaps are 0.10 eV and 0.30 eV, respectively. With reduced $d$, both gaps decrease monotonously till the distance is reduced to 0.315 nm. When the interlayer spacing is further reduced from 0.315 nm to 0.310 nm, the up-spin gap precipitously jumps while the down-spin gap drops. This exceptional sharp change in the energy gap curves is found to be caused by a structural transition from AA stacking to AB stacking near the $C_N$-edge. Near the $C_B$-edge, the AA to AB stacking transition only partially occurs. When the interlayer distance is further reduced, the energy gap of the up-spin channel decreases rapidly while that of the down-spin channel decreases slowly. When the distance is reduced to 0.295 nm, the up-spin gap falls to zero while the down-spin gap is still about 0.13 eV

(Figure 4b). Reduced to this critical distance, the sandwich has undergone a transition from semiconductor to half-metal, with the compressive strain and stress being only about 12% and 1.09 GPa, respectively, which are feasible in current technology. At $d$ = 0.295 nm, the difference electrostatic potential distribution at two edges of ZGNR is hardly changed in comparing with that at $d$ = 0.334 nm, but the relative value is larger as shown in Figure 4c. The charge density redistribution is also greatly enhanced by the coupling effect between the ZBNNR and the ZGNR (Figure 4d).

The physical mechanism of the pressure induced transition in the sandwiched structure is similar to that of the induced transition by a transverse electric field on a ZGNR. Decreasing the interlayer spacing can also enhance the in-built electric field induced by the stronger spontaneous polarization. As interlayer space is reduced, the conduction band bottom of the up-spin channel shifts downward and starts to be filled as a result of the formation of weak interlayer bonds, turning the ZGNR into a magnetic metal. In the same time, the stability of the inter-edge antiferromagnetically coupling in the ZGNR is also attenuated due to the enhanced interlayer interaction, leading to a decrease in the edge magnetic moment. The function of pressure-induced modulation also holds for other Z-BN-G-BNNRs and is robust to any change in the interlayer stacking.

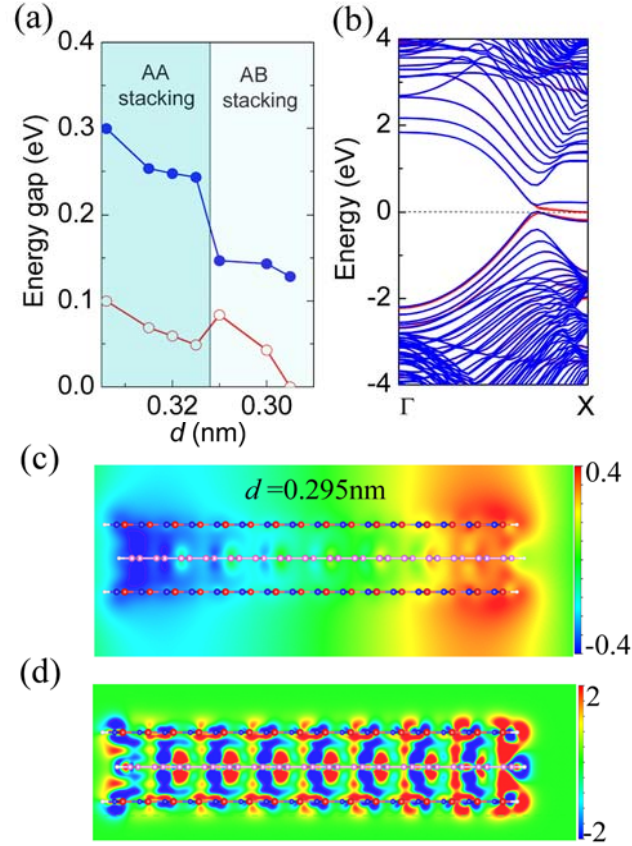

**Figure 4.** (a) Interlayer spacing dependent energy gap of the 16-Z-BN-G-BNNR. The solid and hollow symbols denote the down-spin and up-spin channels, respectively. (b) Band structure of the 16-Z-BN-G-BNNR at the critical interlayer spacing of $d$ = 0.295 nm. The Fermi level is set to 0. The red and blue lines present for up and down spin bands, respectively. (c) The corresponding electrostatic potential difference and (d) charge difference along the ribbon width.

As can be expected, the band gap of such sandwiched structures is also ribbon width dependent. As shown in Figure 5, both the up-spin and down-spin gaps decrease as the ribbon width increases due to the quantum confinement. In the calculated range, the up-spin (down-spin) gaps are always about 0.2 eV lower (higher) than that of the pristine ZGNR. To reflect the accessibility for half-metallicity in the Z-BN-G-BNNRs, we define a rescaled gap as $\delta=|\Delta\alpha-\Delta\beta|/(\Delta\alpha+\Delta\beta)$, where $\Delta\alpha$, $\Delta\beta$ denote for the up-spin gap and down-spin gap of the Z-BN-G-BNNRs, respectively. If $\delta$ is equivalent to 1, the system is an intrinsic half-metal. It is clearly shown in Figure 5 that $\delta$ increases rapidly when the width is narrower than $N_Z=12$, and is over 0.5 for wider ribbons. This means that wider ribbons sandwiched between BN layers should be easier to be tuned into half-metal by applied bias fields.

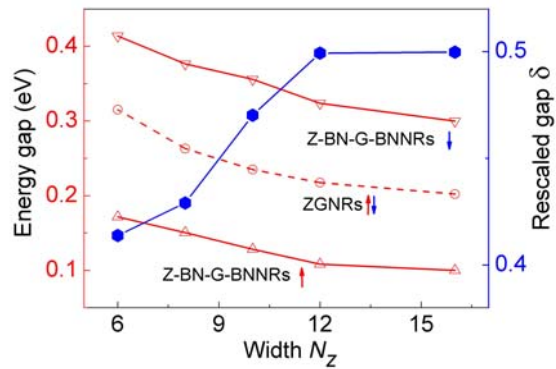

**Figure 5.** Variation of the energy (rescaled) gap with ribbon width. The red dash line and red solid lines are for monolayer ZGNRs and stacked Z-BN-G-BNNRs, respectively. Therein, the arrows denote the spin orientation(s) of each curve. The green dots and line show the rescaled gap against the ribbon width.

In conclusion, we have shown by systematical first-principles calculations that a zigzag GNR sandwiched between two BN nanoribbons or sheets can have novel tunable electronic and magnetic properties due to the strong edge coupling between the BN layers and the ZGNR. This coupling allows a bias-induced spin-resolved band modulation that can endow the system half-metallicity. The half-metallic character can also be achieved by reducing the interlayer spacing of the sandwiched structures. This new insight into the interlayer coupling of BN and graphene nanostructures should have practical importance in pushing GNR into electronics and spintronics applications compatible with the current technology.

**Computational Methods**

Our calculations were performed within the density functional theory as implemented in the VASP code.[25] The projector augmented wave method for the core region and the local spin density approximation (LSDA) for the exchange-correlation potential was used. Test calculations in band structure study using generalized gradient approximation (GGA) including van der Waals interactions gave almost the same results. A kinetic energy cutoff of 530 eV was used in the plane-wave expansion. The Brillouin zone was sampled by 20 special k points for atomic structure relaxation and 60 k points for the electronic structure calculation. The conjugate gradient method was used to optimize the geometry, and all the atoms in the unit cell are fully relaxed until the force on each atom is less than 0.1 eV/nm.

**AUTHOR INFORMATION**

All the corresponding should be addressed to this author: wlguo@nuaa.edu.cn

**ACKNOWLEDGMENT**

This work is supported by the 973 Program (2012CB933403) and National NSF (91023026).

**Supporting Information Available:** The spin-related electronic properties with different stacking configurations. This material is free of charge via the Internet at http://pubs.acs.org.

**REFFERENCES**

(1) Son, Y. W.; Cohen, M. L.; Louie, S. G. Energy Gaps in Graphene Nanoribbons. *Phys. Rev. Lett.* **2006,** *97 (21),* 216803.

(2) Barone, V.; Hod, O.; Gustavo, E. S. Electronic Structure and Stability of Semiconducting Graphene Nanoribbons. *Nano Lett.* **2006,** *6 (12),* 2748-2754.

(3) Jiao, L. Y.; Wang, X. R.; Diankov, G.; Wang, H. L.; Dai, H. J. Facile Synthesis of High-Quality Graphene Nanoribbons. *Nature Nanotechnology.* **2010,** *5,* 321-325.

(4) Obradovic, B.; Kotlyar, R.; Heinz, F.; Matagne, P.; Rakshit, T.; Giles, M. D.; Stettler, M. A.; Nikonov, D. E. Analysis of Graphene Nanoribbons as A Channel Material for Field-Effect Transistors. *Appl. Phys. Lett.* **2006,** *88,* 142102.

(5) Hu, J. N.; Ruan, X. L.; Chen, Y. P. Thermal Conductivity and Thermal Rectification in Graphene Nanoribbons: A Molecular Dynamics Study. *Nano Lett.* **2009,** *9 (7),* 2730–2735.

(6) Guo, Z. X.; Zhang, D. E.; Gong, X. G. Thermal Conductivity of Graphene Nanoribbons. *Appl. Phys. Lett.* **2009,** *95,* 163103.

(7) William, J. E.; Hu, L.; Keblinski, P. Thermal Conductivity of Graphene Ribbons from Equilibrium Molecular Dynamics: Effect of Ribbon Width, Edge Roughness and Hydrogen Termination. *Appl. Phys. Lett.* **2010,** *96,* 203112.

(8) Nakada, K.; Fujita, M.; Dressslhaus, G.; Dresselhaus, M. S. Edge State in Graphene Ribbons: Nanometer Size Effect and Edge Shape Dependence. *Phys. Rev. B* **1996,** *54,* 17954.

(9) Yang, L.; Park, C.; Son, Y. W.; Cohen, M. L.; Louie, S. G. Quasiparticle Energies and Band Gaps in Graphene Nanoribbons. *Phys. Rev. Lett.* **2007,** *99,* 186801.

(10) Pisani, L.; Chan, J. A.; Montanari, B.; Harrison, N. M. Electronic Structure and Magnetic Properties of Graphitic Ribbons. *Phys. Rev. B* **2007,** *75,* 064418.

(11)   Son, Y. W.; Cohen, M. L.; Louie, S. G. Half-metallic Graphene Nanoribbons. *Nature* **2006,** *444,* 347-349.

(12)   Kan, E. J.; Li, Z. Y.; Yang, J. L.; Hou, J. G. Will Zigzag Graphene Nanoribbon Turn to Half Metal under Electric Field? *Applied Phys. Lett.* **2007,** *91,*243116.

(13)   Guo,Y. F.; Guo, W. L. Semiconducting to Half-Metallic to Metallic Transition on Spin-Resolved Zigzag Bilayer Graphene Nanoribbons. *J. Phys. Chem. C* **2010,** *114,* 13098.

(14)   Hod, O.; Barone, V.; Peralta, J. E.; Scuseria, G. E. Enhanced Half-Metallicity in Edge-Oxidized Zigzag Graphene Nanoribbons. *Nano. Lett.* **2007,** *7 (8),* 2295–2299

(15)   Kan, E. J.; Wu, X. J.; Li, Z. Y.; Zeng, X. C.; Yang, J. L.; Hou, J. G. Half-metallicity in Hybrid BCN Nanoribbons. *J. Chem. Phys.* **2008**, *129 (8),* 084712.

(16)   Ding, Y.; Wang, Y. L.; Ni, J. Electronic Properties of Graphene Nanoribbons Embedded in Boron Nitride Sheets. *Appl. Phys. Lett.* **2009,** *95 (12),* 123105.

(17)   Zeng, H. B.; Zhi, C. Y.; Zhang, Z. H.; Wei, X. L.; Wang, X. B.; Guo, W. L.; Bando, Y.; Golberg, D. "White Graphenes": Boron Nitride Nanoribbons via Boron Nitride Nanotube Unwrapping. *Nano Lett.* **2010,** *10 (12),* 5049–5055.

(18)   Zhang, Z. H.; Guo, W. L. Energy-Gap Modulation of BN Ribbons by Transverse Electric Fields: First-Principles Calculations. *Phys. Rev. B* **2008,** *77 (7),* 075403.

(19)   Sławińska, J.; Zasada, I.; Kosiński, P.; Klusek, Z. Reversible Modifications of Linear Dispersion: Graphene between Boron Nitride Monolayers. *Phys. Rev. B* **2010,** *82 (8),* 085431.

(20)   Sakai, Y.; Koretsune, T.; Saito, S. Electronic Structure and Stability of Layered Superlattice Composed of Graphene and Boron Nitride Monolayer. *Phys. Rev. B* **2011,** *83 (20),* 205434.

(21)   Kan, M.; Zhou, J.; Wang, Q.; Sun, Q.; Jena, P. Tuning the Band Gap and Magnetic Properties of BN Sheets Impregnated with Graphene Flakes. *Phys. Rev. B* **2011,** *84 (20),* 205412.

(22)  Quhe, R.; Zheng, J. X.; Luo, G. F.; Liu, Q. H.; Qin, R.; Zhou, J.; Yu, D. P.; Nagase, S.; Mei, W. N.; Gao, Z. X. *et al.* Tunable and Sizable Band Gap of Single-Layer Graphene Sandwiched between Hexagonal Boron Nitride. *NPG Asia Mater.* **2012,** *4,* e6.

(23)  Csányi, G.; Littlewood, P. B.; Nevidomsky, A. H.; Pickard, C. J.; Simons, B. D. The Role of The Interlayer State in The Electronic Structure of Superconducting Graphite Intercalated Compounds. *Nature Phys.* **2005**, *1,* 42-45.

(24)  Landman, U.; Luedtke, W. D.; Nancy, A. B.; Richard, J. C. Atomistic Mechanisms and Dynamics of Adhesion, Nanoindentation and Fracture. Science **1990,** 248 (4954), 454-461.

(25)  Kresse, G.; Hafner, J. Ab Initio Molecular Dynamics for Liquid Metals. *Phys. Rev. B* **1993**, *47,* 558; Kresse, G.; Hafner, J. Ab Initio Molecular Dynamics Simulation of The Liquid-Metal–Amorphous-Semiconductor Transition in Germanium. *Phys. Rev. B* **1994**, *49,* 14251.